\documentstyle{article}  
\input{amssym.def} \input{amssym.tex}
\newtheorem{Definition}{Definition} \newtheorem{Lemma}{Lemma}
\newtheorem{Theorem}{Theorem} \newtheorem{Proposition}{Proposition}
\newcommand{\Ad}{{\rm Ad}}
\newcommand{\bg}{{\bf g}} 

\renewcommand{\ll}{\label} \newcommand{\Ah}{{\frak A}^{\hbar}}
\newcommand{\be}{\begin{equation}} \newcommand{\ee}{\end{equation}}
\newcommand{\bea}{\begin{eqnarray}} \newcommand{\eea}{\end{eqnarray}}
 \newcommand{\bib}{\bibitem}
\newcommand{\ci}{\cite} \newcommand{\ca}{$C^*$-algebra}
\newcommand{\rep}{representation} \newcommand{\irrep}{irreducible
representation} 
 \newcommand{\ovl}{\overline}
 \newcommand{\til}{\tilde}
\newcommand{\raw}{\rightarrow} 
\newcommand{\n}{\parallel} 
 
\newcommand{\x}{\times}  \newcommand{\Co}{{\rm
Co}} \newcommand{\cin}{C^{\infty}} \newcommand{\cci}{C^{\infty}_c}
\newcommand{\q}{{\cal Q}_{\hbar}}
\newcommand{\lho}{\lim_{\hbar\rightarrow 0}} \newcommand{\inv}{^{-1}}
\newcommand{\Exp}{{\rm Exp}} \newcommand{\CPW}{C^{\infty}_{\mbox{\tiny
PW}}}  
\newcommand{\gm}{\gamma} \newcommand{\Gm}{\Gamma}
\newcommand{\dl}{\delta}

\newcommand{\th}{\theta} 
 \newcommand{\kp}{\kappa}

\newcommand{\ch}{\chi}  
\newcommand{\om}{\omega} 
\newcommand{\A}{{\frak A}} \newcommand{\GC}{{\frak C}}
\newcommand{\g}{{\frak g}}

\newcommand{\CN}{{\cal N}} \newcommand{\CS}{{\cal S}}
\newcommand{\CQ}{{\cal Q}} 
 \newcommand{\R}{{\Bbb R}}
\newcommand{\T}{{\Bbb T}} \newcommand{\Z}{{\Bbb Z}} \makeatletter
\newskip\tempskip \def\endproof{{\parfillskip24\p@ plus\@ne
fil\@@par}\tempskip\prevdepth
\ifdim\lastskip=\z@\tempskip\z@\else\vskip-\lastskip
\ifdim\tempskip>4\p@ \tempskip.5\tempskip \else \tempskip\z@\fi\fi
\nobreak\vskip-\baselineskip\vskip-\tempskip\noindent\hbox
to\hsize{\hfill
$\blacksquare$}\par\vskip\tempskip\vskip\abovedisplayskip\@doendpe}
\makeatother

\newcommand{\enp}{\endproof}
\topmargin = - 0.5 cm
\begin{document} 
\setlength{\baselineskip}{1\baselineskip}
\thispagestyle{empty} \setlength{\unitlength}{1cm} \title{Twisted Lie
group $C^*$-algebras as strict quantizations}
\author{N.P.~Landsman\thanks{Supported by a fellowship from the Royal
Netherlands Academy of Arts and Sciences (KNAW)}\\ \mbox{}\hfill \\
Korteweg-de Vries Institute for Mathematics \\ University of Amsterdam
\\ Plantage Muidergracht 24 \\ 1018 TV AMSTERDAM, THE NETHERLANDS \\
\mbox{}\hfill \\ {\em email:} npl@wins.uva.nl } \date{\today}
\maketitle
\begin{abstract}
A nonzero 2-cocycle $\Gm\in Z^2(\g,\R)$ on the Lie algebra $\g$ of a
 compact Lie group $G$ defines a twisted version of the Lie-Poisson
 structure on the dual Lie algebra $\g^*$, leading to a Poisson
 algebra $\cin(\g_{(\Gm)}^*)$.  Similarly, a multiplier $c\in
 Z^2(G,U(1))$ on $G$ which is smooth near the identity defines a twist
 in the convolution product on $G$, encoded by the twisted group \ca\
 $C^*(G,c)$.

Further to some superficial yet enlightening analogies between
 $\cin(\g^*_{(\Gm)})$ and $C^*(G,c)$, it is shown that the latter is a
 strict quantization of the former, where Planck's constant $\hbar$
 assumes values in $(\Z\backslash\{0\})\inv$. This means that there
 exists a continuous field of \ca s, indexed by $\hbar\in 0\cup
 (\Z\backslash\{0\})\inv$, for which $\A^0=C_0(\g^*)$ and
 $\Ah=C^*(G,c)$ for $\hbar\neq 0$, along with a cross-section of the
 field satisfying Dirac's condition asymptotically relating the
 commutator in $\Ah$ to the Poisson bracket on $\cin(\g^*_{(\Gm)})$.
 Note that the `quantization' of $\hbar$ does not occur for $\Gm=0$.
 \end{abstract}\thispagestyle{empty} \newpage
\section{Introduction}
 There now exists a satisfying $C^*$-algebraic definition of
quantization, which enables one to link Poisson and symplectic
geometry with operator algebras and non-commutative geometry. The main
functional-analytic idea behind this goes back to Rieffel \ci{Rie1},
who showed how the idea of `formal' deformation quantization \ci{Bay}
may be adapted to an operator-algebraic context. Later modifications
by Rieffel himself and by the author have culminated in the following
definition (see \ci{MT} for references and comments).  Recall that
$C_0(P)$ is the commutative \ca\ of continuous functions on $P$ which
vanish at infinity, equipped with the supremum-norm.
\begin{Definition}\ll{defqua}  A continuous
quantization of a Poisson manifold $P$ consists of a subset
$I\subseteq\R$ (containing $0$ as an accumulation point), a continuous
field of \ca s $(\{\Ah\}_{\hbar\in I},\GC\subset\prod_{\hbar\in
I}\Ah)$ with $\A^0=C_0(P)$, a Poisson algebra $\til{\A}^0$ which lies
densely in $C_0(P)$, and a distinguished collection $\{\CQ(f)\}_{f\in
\til{\A}^0}\subset \GC$ of cross-sections, such that \bea \CQ_0(f)& =
& f; \ll{q0f}\\ \q(f^*) & = & \q(f)^*;\ll{real} \eea for all $\hbar\in
I$ and $f\in \til{\A}^0$. Finally, Dirac's condition \be \lho \n
\frac{i}{\hbar}[\q(f),\q(g)] -\q(\{f,g\})\n =0 \ll{direq} \ee should
hold for all $f,g\in \til{\A}^0$.
\end{Definition}

We refer to Dixmier \ci{Dix} for the concept of a continuous field of
\ca s; the collection $\GC$ of cross-sections determines a continuity
structure on $\{\Ah\}_{\hbar\in I}$, and has to satisfy a number of
conditions which are listed in \ci{Dix}. It should be noted that
Definition \ref{defqua} guarantees the property \be \lho \n \q(f)
\q(g)-\q(f g)\n =0 \ll{cheq} \ee for all $f,g\in\til{\A}^0$. In
addition, the function $\hbar\raw \n\q(f)\n$ is continuous on $I$, so
that, in particular, one has \be \lho \n \q(f)\n =\n f\n_{\infty},
\ll{bohreq} \ee where the right-hand side is the supremum-norm of $f$.
Conversely, one has
\begin{Lemma}\ll{sdgcfca}
Suppose one has a Poisson manifold $P$, a family $\{\Ah\}_{\hbar\in
I}$ of \ca s indexed by a discrete subset $I\subset\R$ containing $0$
as an accumulation point, a Poisson algebra $\til{\A}^0$ whose
(sup-norm) closure is $\A^0=C_0(P)$, and a collection of linear maps
$\{\q:\til{\A}^0\raw \Ah\}_{\hbar\in I}$ satisfying (\ref{q0f}),
(\ref{real}), (\ref{direq}), (\ref{cheq}), and (\ref{bohreq}).

There exists a family $\GC\subset\prod_{\hbar\in I}$ making
$(\{\Ah\}_{\hbar\in I},\GC)$ into a continuous field of \ca s, such
that $\GC$ contains all maps $\{\q(f)\}_{\hbar\in I}$,
$f\in\til{\A}^0$.
\end{Lemma}
The continuous field in question is uniquely determined when the set
$\{\q(f)\}_{f\in \til{\A}^0}$ is dense in $\Ah$ for all $\hbar\in I$,
but we shall not need this. A proof of this lemma may be found in
\ci{MT}.

Apart from its more stringent definition of convergence, strict
quantization as defined by Definition \ref{defqua} differs from
deformation quantization in the sense of \ci{Bay}, as well as from the
corresponding notion in \ci{Rie1}, in that $\q(\til{\A}^0)$ is not
necessarily closed under multiplication (in $\A^{\hbar}$). If it is,
and if $\q$ is non-degenerate in that $\q(f)=0$ iff $f=0$ for each
$\hbar$, one may define an associative `deformed' product
$\cdot_{\hbar}$ in $\til{\A}^0$ with the property
$\q(f)\q(g)=\q(f\cdot_{\hbar}g)$ (and, of course, $f\cdot_0 g=fg$).
The conditions on a strict quantization may then be rephrased in terms
of this product in the obvious way, leading to the framework of
\ci{Rie1}. However, there are many examples of strict quantization
that are not deformation quantizations, including the ones in this
paper that correspond to nontrivial group extensions.

A conceptually rather pleasing class of examples of strict
quantization (in the original definition of \ci{Rie1}) was discovered
by Rieffel \ci{Rie2}.  Consider a Lie group $G$ with Lie algebra
$\g$. The dual $\g^*$ of $\g$ is a Poisson manifold under the
well-known Lie-Poisson bracket \ci{LM} \be \{f,g\}(\th)= -
\th([df_{\th},dg_{\th}]). \ll{liepbr} \ee

The symplectic leaves of a Poisson manifold $P$ (along with their
covering spaces) play the role of `classical' irreducible \rep s of
the corresponding Poisson algebra $\cin(P)$ \ci{MT}.  As shown by
Kirillov \ci{Kir} (also cf.\ \ci{LM}), the symplectic leaves of $\g^*$
with respect to this Poisson structure are (the connected components
of) its coadjoint orbits. This result is reminiscent of the bijective
correspondence between the (non-degenerate) \irrep s of the group \ca\
$C^*(G)$ and the irreducible unitary \rep s of $G$ \ci{Ped}.  Since
the latter may be seen as the quantum counterparts of the coadjoint
orbits of $G$, Kirillov's result already suggests that $C^*(G)$ should
be the \ca ic analogue of the Poisson algebra $\cin(\g^*)$.

The correspondence between the coadjoint orbits in $\g^*$ and the
unitary \irrep s of $G$ is at its best (namely, bijective and
functorial) when $G$ is nilpotent, connected, and simply connected.
In that case, Rieffel \ci{Rie2} showed that $C^*(G)$ is related to
$\cin(\g^*)$ by a strict quantization, with $I=\R$. Under the stated
assumptions $G$ is exponential, so that one may identify $G$ with
$\g$.  Translated into the setting of Definition \ref{defqua}, the
quantization maps $\q$ are given by \be
\q(f):X\raw\int_{\g^*}\frac{d^n\th}{(2\pi\hbar)^n}\,
e^{i\th(X)/\hbar}f(\th). \ll{defqhc} \ee Here
$f\in\til{\A}^0=\CS(\g^*)$, the Schwartz space of test functions on
$\g^*$.

In section 2 we show that an analogous statement holds for arbitrary
compact Lie groups; given the results on strict quantization on
Riemannian manifolds in \ci{Lan}, this is a simple exercise.  In
section 3 we modify the Lie-Poisson structure on $\g^*$ by a nonzero
2-cocycle $\Gm$ on $\g$, and show that $C^*(G)$ should then be
replaced by the twisted group \ca\ $C^*(G,c)$, defined by a multiplier
$c$ on $G$ which is smooth near the identity.  Similar \rep-theoretic
analogies as in the untwisted case then hold.  The main point of this
Letter is made in section 4, where we extend the strict quantization
of the Lie-Poisson structure to the twisted case.  That is, when $\Gm$
is the derivative of $c$ in a suitable sense, we show that the \ca\
$C^*(G,c)$ and the twisted Poisson algebra $\cin(\g^*_{(\Gm)})$ are
related by a strict quantization. Surprisingly, this only works if the
interval $I=\R$ is replaced by the discrete set $I=0\cup
(\Z\backslash\{0\})\inv$.  The quantization maps, however, are still
given by (\ref{defqhc}).
\section{Strict quantization of the Lie-Poisson structure for compact
Lie groups} In this section we modify (\ref{defqhc}) so as to make it
applicable to compact Lie groups. Firstly, the Fourier transform of
$f\in L^1(\g^*)$ is defined by \be
\grave{f}(X)=\int_{\g^*}\frac{d^n\th}{(2\pi)^n}\, e^{i\th(X)}f(\th),
\ll{ftongst} \ee where $d^n\th$ is Lebesgue measure on
$\g^*\simeq\R^n$, whose normalization is fixed by that of the Haar
measure $dx$ on $G$, as follows. When $f$ has support near $e$, we can
write $ \int_G dx\, f(x)=\int_{\g} d^nX \, J(X)f(\Exp(X))$, where
$d^nX$ is a Lebesgue measure on $\g$, and $J$ is some Jacobian. The
normalization is now fixed by the condition $J(0)=1$. In turn, the
normalization of the Lebesgue measure $d^n\th$ on $\g^*$ is fixed by
requiring the inversion formula $f(\th)=\int_{\g} d^n X\,
e^{-i\th(X)}\grave{f}(X)$.  We define $\CPW(\g^*)$ as the class of
functions on $\g^*$ whose Fourier transform $\grave{f}$ is in
$\cci(\g)$. This is a Poisson subalgebra of $\cin(\g^*_{\pm})$.

We choose a smooth cutoff function $\kp$ on $\g$ which equals $1$ in a
neighbourhood $\til{{\cal N}}$ of $0$, is invariant under inversion
$X\raw -X$, and has support in the neighbourhood $\cal N$ of $0$ on
which $\Exp$ is a diffeomorphism. When $G$ is compact one may assume
that $\kp$ is $\Ad$-invariant, i.e., satisfies $\kp(\Ad(y)X)=\kp(X)$
for all $y\in G$.  This may always be achieved by averaging.

We now modify (\ref{defqhc}) as follows: for $x\notin\Exp({\cal N})$
we put $\q(f)(x)=0$, whereas for $x\in\Exp({\cal N})$ we put \be
\q(f)(x)=\hbar^{-n}
\kp(\Exp\inv(x))\grave{f}(\Exp\inv(x)/\hbar).\ll{defqhc2} \ee

The restriction $f\in\CPW(\g^*)$ implies that for small enough $\hbar$
the operator $\q(f)$ is independent of $\kp$.
\begin{Theorem}\ll{defgG}
Suppose $G$ is an $n$-dimensional compact Lie group.  The collection
of maps $\q:\CPW(\g^*) \raw C^*(G)$ defined by (\ref{defqhc2}) and
preceding text, where $\hbar\in \R\backslash\{0\}$, satisfies
(\ref{q0f}), (\ref{real}), (\ref{direq}), (\ref{cheq}), and
(\ref{bohreq}). Hence there exists a strict quantization (cf.\
Definition \ref{defqua}) of $\g^*$ on $I=\R$ for which
$\til{\A}^0=\CPW(\g^*)$, $\A^0=C_0(\g^*)$, and $\Ah=C^*(G)$ for
$\hbar\notin 0$, the maps $\q(f)$ being cross-sections of the
associated continuous field of \ca s.
\end{Theorem}

The conclusion of the theorem is immediate from Lemma \ref{sdgcfca}.
To prove that the assumptions of the lemma are satisfied, we identify
$C^*(G)$ with $\pi_L(C^*(G))$, where $\pi_L$ is the left-regular \rep\
on $L^2(G)$ \ci{Ped}; this \rep\ is faithful because compact groups
are amenable.  Also, we identify $\cin(\g^*)$ as a Poisson algebra
with the subalgebra $\cin(T^*G)^R$ of right-invariant smooth functions
on $T^*G$, equipped with the canonical cotangent bundle Poisson
bracket \ci{LM}.  This identification is inherited by
$\CPW(\g^*)\simeq \CPW(T^*G)^R$, where on the right-hand side the
class $\CPW$ is defined relative to the Fourier transform in the fiber
direction \ci{Lan,LM}; recall that $T^*_xG\simeq\g^*$.

A compact Lie group $G$ admits a right-invariant Riemannian metric
$\bg$, such that the exponential map $\exp_e$ obtained from $\bg$
coincides with the map $\Exp$ defined by the Lie group structure
\ci{Mil}. Using such a metric, the generalized Weyl quantization
prescription on Riemannian manifolds of \ci{Lan}, restricted to
$\CPW(T^*G)^R$, coincides with $\q$ as defined by (\ref{defqhc2}).
All claims then follow from Theorem 1 in \ci{Lan}.  \enp
\section{The twisted Lie-Poisson algebra vs the twisted group algebra}
Let $\Gm\in Z^2(\g,\R)$ be a 2-cocycle on $\g$ with values in $\R$
\ci{LM}.  This leads to a modification of the Lie-Poisson structure on
$\g^*$, in which one adds a term $-\Gm(df,dg)$ to the right-hand side
of (\ref{liepbr}). In canonical co-ordinates on $\g^*$ (relative to a
basis $\{T_a\}_{a=1,\ldots,n}$ of $\g$), the ensuing bracket reads \be
\{f,g\}^{(\Gm)}_{\pm}=- \left( C_{ab}^c\th_c+\Gm_{ab}\right)
\frac{\partial f}{\partial \th_a} \frac{\partial g}{\partial \th_b} ,
\ll{modliepo2} \ee where the $C_{ab}^c$ are the structure constants of
$\g$ in the given basis, and $\Gm_{ab}=\Gm(T_a,T_b)$.  We denote the
space $\g^*$, seen as a Poisson manifold through (\ref{modliepo2}), by
$\g^*_{(\Gm)}$, with associated Poisson algebra $\cin(\g^*_{(\Gm)})$.

The 2-cocycle $\Gm$ defines a central extension $\g_{\Gm}$ of $\g$ as
well. As a vector space one has $\g_{\Gm}=\g\oplus\R$; denoting the
central element by $T_0$ (this is a basis vector in the extension
$\R$), the new Lie bracket is $[X,Y]_{\Gm} = [X,Y]+\Gm(X,Y)T_0$. This
also equips the dual $\g_{\Gm}^*$ with the Lie-Poisson structure.  Let
$\om^0$ be the basis element in $\g^*_{\Gm}$ dual to $T_0$. Then
$J_1:\g^*_{(\Gm)}\raw \g_{\Gm}$ given by $J_1(\th)=\th+\om^0$ (where
$\g^*$ is embedded in $\g^*_{\Gm}$ as the annihilator of the extension
$\R$) is a Poisson map.
\begin{Proposition}\ll{cggmcgmg}
The canonical identification of $\cin(\g^*_{\Gm})/\ker(J_1^*)$ with
$\cin(\g^*_{(\Gm)})$ is a Poisson isomorphism.
\end{Proposition}
  
This is immediate from the definitions and (\ref{modliepo2}).  \enp

The identification between the symplectic leaves in $\g^*$ with
respect to the Lie-Poison structure and the coadjoint orbits has the
following generalization to the twisted case \ci{LM,MT}.  Let $\gm$ be
a symplectic cocycle on $G$ with the property that $\Gm(X,Y) =-
\frac{d}{dt}\gm(\Exp(tX))(Y)_{|t=0}$.  The symplectic leaves of
$\g^*_{(\Gm)}$ then coincide with the $G$-orbits in $\g^*$ under the
twisted coadjoint action $\Co^{\gm}(x)\th=\Co(x)\th+\gm(x)$, where
$\Co$ stands for the usual coadjoint action.

We pass from Poisson algebras to \ca s. The role of $\Gm$ is now
played by a multiplier $c\in Z^2(G,U(1))$ on $G$ which is smooth near
the identity \ci{TW,MT}.  The quantum analogue of the twisted Poisson
algebra $\cin(\g^*_{(\Gm)})$ is the twisted group algebra
$C^*(G,c)$. This is defined as a suitable $C^*$-completion of
$L^1(G)$, under the twisted convolution product \be f*g(x)=\int_G dy\,
c(xy\inv,y) f(xy\inv)g(y), \ll{defconv} \ee and the twisted involution
\be f^*(x)=\ovl{c(x,x\inv)f(x\inv)}. \ll{definvga} \ee

The bijective correspondence between the non-degenerate (irreducible)
\rep s of $C^*(G)$ and the continuous unitary (irreducible) \rep s of
$G$ is generalized to a bijective correspondence between the
non-degenerate (irreducible) \rep s of $C^*(G,c)$ and the continuous
projective unitary (irreducible) \rep s of $G$ with multiplier $c$;
see \ci{BS,MT}.

Furthermore, a multiplier $c$ defines a central extension $G_c$ of $G$
by $U(1)$ \ci{TW}. A quantum analogue of Proposition \ref{cggmcgmg} is
as follows.
\begin{Proposition}\ll{ninZ}
Let $G$ be a compact Lie group with multiplier $c$, and write $\pi^k$
for the \rep\ of $C^*(G_c)$ corresponding to the \rep\ $U^k(G_c)$
induced by $U_k(U(1))$, where $k\in\Z$ and $U_k(z)=z^k$ for $z\in \Bbb
T =U(1)$.  For each $k\in\Z$ there are isomorphisms \be
C^*(G,c^k)\simeq \pi^k(C^*(G_c))\simeq
C^*(G_c)/\ker(\pi^k). \ll{cstcnrsim} \ee Explicitly, under the first
isomorphism the function $\pi^k(f)\in C^*(G,c^k)$ is \be
\pi^k(f):x\raw \int_{\T} dz\, z^k f(x,z). \ll{pinf} \ee
\end{Proposition}

Here $dz$ is the normalized Haar measure on $\T$.  Given a projective
\rep\ $U(G)$ with multiplier $c^k$, one defines an associated \rep\
$U_{c^k}$ of $G_c$ by $U_{c^k}(x,z)=z^kU(x)$, and verifies that
$U_{c^k}$ is unitarily equivalent to the \rep\ $U^k(G_c)$ induced by
$U_k(U(1))$.\enp

This proposition is closely related to the decomposition \be C^*(G_c)
\simeq \oplus_{k\in\Z} \, \pi^k(C^*(G_c)), \ll{decpilcgc} \ee which
follows from the isomorphism $C^*(G_c)= \pi_L(C^*(G_c))$ and the
Peter-Weyl theorem applied to $G_c$.
\section{Strict quantization of the twisted Lie-Poisson structure for compact
Lie groups} Comparing the comment after the proof of Proposition
\ref{cggmcgmg} with the one following (\ref{definvga}), and also
comparing Propositions \ref{cggmcgmg} and \ref{ninZ}, it is clear that
the twisted group \ca\ $C^*(G,c)$ is indeed a quantum version of the
twisted Poisson algebra $\cin(\g^*_{(\Gm)})$.  Inspired by the
analogies in question, we now generalize Theorem \ref{defgG} to the
twisted case.

 We identify $\R$ in $\g_{\Gm}=\g\oplus\R$ with the Lie algebra
${\frak u}_c(1)$ of the central subgroup $U(1)\subset G_c$ defining
the extension, and write $\Exp:{\frak u}(1)\raw U(1)$ for the
exponential map, conventionally realized as $\Exp(X)=\exp(-iX)$.  In a
neighbourhood $\CN_e\x\CN_e$ of $(e,e)$ we can write $c=\Exp(\ch)$,
where $\ch:\CN_e\x \CN_e\raw {\frak u}_c(1)$. Then define
$\Gm:\g\x\g\raw\R$ by \be \Gm(X,Y) = \frac{d}{ds}\frac{d}{dt}\left
[ \ch(\Exp(tX),\Exp(sY))-\ch(\Exp(sY),\Exp(tX))
\right]_{|s=t=0}. \ll{Gmc} \ll{Gmfrompr} \ee It is easy to see that
$\Gm\in Z^2(\g,\R)$ when $c\in Z^2(G,U(1))$.
\begin{Theorem}\ll{cgGmtoccg}
Suppose $G$ is an $n$-dimensional compact Lie group, with multiplier
$c\in Z^2(G,U(1))$, and define a 2-cocycle $\Gm\in Z^2(\g,\R)$ on $\g$
by (\ref{Gmc}).  Regard $\CPW(\g_{(\Gm)}^*)$ as a Poisson subalgebra
of $\cin(\g_{(\Gm)}^*)$ with respect to the Poisson bracket
(\ref{modliepo2}), and regard (\ref{defqhc2}) as a map from
$\CPW(\g_{(\Gm)}^*)$ to $\Ah=C^*(G,c^{1/\hbar})$, where $\hbar\in
(\Z\backslash\{0\})\inv$.

 The collection of maps $\q$, thus construed, satisfies (\ref{q0f}),
 (\ref{real}), (\ref{direq}), (\ref{cheq}), and (\ref{bohreq}). Hence
 there exists a strict quantization (cf.\ Definition \ref{defqua}) of
 $\g_{(\Gm)}^*$ for which $\til{\A}^0=\CPW(\g_{(\Gm)}^*)$,
 $\A^0=C_0(\g_{(\Gm)}^*)$, and $\Ah=C^*(G,c^{1/\hbar})$ for $\hbar\in
 (\Z\backslash\{0\})\inv$, the maps $\q(f)$ being cross-sections of
 the associated continuous field of \ca s.
\end{Theorem}

It is obvious that (\ref{real}) holds.  The proof of the other
properties is based on the analogy between Propositions \ref{cggmcgmg}
and \ref{ninZ}.  Extend $f\in \CPW(\g^*)$ to a function $\til{f}\in
\CPW(\g_{\Gm}^*)$, such that $f(\th)=\til{f}(1,\th)$ and \be
\til{f}(\th_0\neq 1,\th)<\til{f}(\th_0=1,\th)=f(\th); \ll{ineqftil}
\ee in particular, one has \be \n f\n_{\infty}=\n
\til{f}\n_{\infty}. \ll{normftil} \ee In view of (\ref{modliepo2})
this automatically means that \be
\{\til{f},\til{g}\}(1,\th)=\{f,g\}^{\Gm}(\th), \ll{equalpb} \ee since
the left-hand side does not involve derivatives with respect to
$\th_0$.

We denote (\ref{defqhc2}) as defined on $\CPW(\g_{(\Gm)}^*)$, taking
values in $C^*(G,c^{1/\hbar})$, by $\q$, whereas the map defined in
the same way, but now on $\CPW(\g_{\Gm}^*)$, taking values in
$C^*(G_c)$, is written as $\til{\CQ}_{\hbar}$.  A short computation
using (\ref{pinf}) and an elementary oscillatory integral shows that
\be \pi^{1/\hbar}(\til{\CQ}_{\hbar}(\til{f}))=\q(f) \ll{proofcgGm} \ee
for $\hbar\in (\Z\backslash\{0\})\inv$ small enough so that the
right-hand side is independent of $\kp$.  In particular, the left-hand
side only depends on the value of $\til{f}$ at $\th_0=1$; this is a
special case of the fact that, for $\hbar$ small enough,
$\pi^{k}(\til{\CQ}_{\hbar}(\til{f}))$ only depends on
$\til{f}(\th_0=k\hbar)$. This follows by a similar calculation as the
one leading to (\ref{proofcgGm}).

Theorem \ref{defgG} applied to $G_c$ implies that $\lho \n
\til{\CQ}_{\hbar}(\til{f})\n =\n\til{f}\n_{\infty}$. On the other
hand, according to (\ref{decpilcgc}) one has $\n A\n=\sup_{k\in\Z}
\n\pi^k(A)\n$ for all $A\in C^*(G_c)$. Combining the two of these
equations with the last remark of the preceding paragraph and the
property (\ref{ineqftil}), we conclude that \be \lho \n
\til{\CQ}_{\hbar}(\til{f})\n = \lho \n
\pi^{1/\hbar}(\til{\CQ}_{\hbar}(\til{f}))\n=\n\til{f}\n_{\infty}.  \ee
Together with (\ref{normftil}) and (\ref{proofcgGm}) this proves
(\ref{bohreq}).

Equations (\ref{cheq}) and (\ref{direq}) now follow from
(\ref{proofcgGm}), Proposition \ref{ninZ}, (\ref{equalpb}), Theorem
\ref{defgG} (once again applied to $G_c$), and the inequality $\n
\pi^k(A)\n\, \leq\,\n A\n$ in $C^*(G_c)$.  \enp

While proved for compact $G$, Theorem \ref{cgGmtoccg} may hold in
other situations.  For example, let $G\simeq \g^*=\R^{2n}$, with $\Gm$
given by $\Gm(P_i,P_j) = \Gm(Q^i,Q^j)=0$ and $\Gm(P_i,Q^j) =
-\dl_i^j$, and $c$ defined by $c((u,v),(u',v'))=e^{i(uv'-vu')/2}$.
Then the statement of Theorem \ref{cgGmtoccg} holds as well.

\end{document}